\documentclass[10pt, conference, onecolumn, compsocconf]{IEEEtran}

\usepackage[small]{caption2}
\usepackage[ruled,noend,linesnumbered]{algorithm2e}
\usepackage{graphicx,color}
\usepackage{amssymb}
\usepackage{amsmath}
\usepackage{multicol}

\begin{document}

\title{Improving PPM Algorithm Using Dictionaries}

\author{
\IEEEauthorblockN{Yichuan Hu}
\IEEEauthorblockA{Department of Electrical and Systems Engineering\\
University of Pennsylvania\\
Email: yichuan@seas.upenn.edu}
\and
\IEEEauthorblockN{Jianzhong (Charlie) Zhang, Farooq Khan and Ying Li}
\IEEEauthorblockA{Standards Research Laboratory\\
Samsung Telecommunications America\\
Email: \{jzhang,fkhan,yli2@sta.samsung.com\}}
}
\maketitle

\begin{abstract}
We propose a method to improve traditional character-based PPM text compression algorithms. Consider a text file as a sequence of alternating words and non-words, the basic idea of our algorithm is to encode non-words and prefixes of words using character-based context models and encode suffixes of words using dictionary models. By using dictionary models, the algorithm can encode multiple characters as a whole, and thus enhance the compression efficiency. The advantages of the proposed algorithm are: 1) it does not require any text preprocessing; 2) it does not need any explicit codeword to identify switch between context and dictionary models; 3) it can be applied to any character-based PPM algorithms without incurring much additional computational cost. Test results show that significant improvements can be obtained over character-based PPM, especially in low order cases. \footnote{Work presented in this report is done by Yichuan Hu during his intern at Standards Research Lab, Samsung Telecommunications America in 2009. This report is a longer version of the paper published in 2011 Data Compression Conference, Snowbird, UT \cite{hu_dcc11}.}
\end{abstract}


\section{Introduction}

Prediction by partial matching (PPM) \cite{cleary_tcom84} has set a benchmark for text compression algorithms due to its high compression efficiency. In PPM, texts are modeled as Markov processes in which the occurrence of a character only depends on its context, i.e., $n$ preceding characters, where $n$ is called the context order and is a parameter of PPM. In each context, probabilities of next character is maintained. When a new character comes, its probability is estimated using context models and is encoded by an arithmetic coder. During the encoding/decoding process, the context models (probability tables) are updated after a character is encoded/decoded. By using such adaptive context models, PPM is able to predict text as well as human do \cite{teahan_dcc96}, and achieves higher compression ratio than other compression algorithms \cite{bell_89}.

However, one limitation of PPM is that the prediction is character-based, i.e., characters are encoded one by one sequentially, which is not quite efficient. In \cite{moffat_word}, word-based PPM is proposed in which every word is predicted by its preceding words, but its performance is even worse than character-based PPM \cite{bell_89}, mainly because the alphabet size of English words is so large that very long texts are required to collect sufficient statistical information for word-based models. Similarly, Horspool \cite{horspool_dcc92} introduced an algorithm that alternatively switches between word-based and character-based PPM, but it needs to explicitly encode the length of characters when character-based PPM is used, resulting in unnecessary overhead. Recently, Skibi\'{n}ski \cite{skibinski_06} extended the alphabet of PPM to long repeated strings by pre-processing the whole texts before encoding, which showed performance improvement over traditional PPM.

In this paper, we propose an enhanced algorithm that combines traditional character-based context models and dictionary models. The basic idea is that, for most English words, given the first a few characters (prefix) the rest of the word (suffix) can be well predicted. Specifically, in addition to context models for character prediction used in traditional PPM \cite{cleary_tcom84} we introduce dictionary models that contain words with a common prefix. By doing so, a word can be predicted and encoded given its prefix, i.e., the first a few characters. Therefore, different from traditional PPM \cite{cleary_tcom84} and its variations \cite{moffat_tcom90,cleary_unbounded,bloom_ppmz,shkarin_dcc02}, proposed algorithm can achieve variable-length prediction by partial matching (VLPPM). The remainder of the paper is organized as follows. Section \ref{sec_vlp} describes the dictionary model used for variable-length prediction and the framework of proposed scheme. Section \ref{sec_details} details the encoding and decoding algorithms. Test results and conclusions are presented in Section \ref{sec_results} and \ref{sec_con}, respectively.

\section{Variable-Length Prediction}
\label{sec_vlp}

\subsection{Dictionary Model For Variable-Length Prediction}

\begin{figure}[t]
   \begin{center}
   \includegraphics[width=0.6\columnwidth,keepaspectratio,clip]{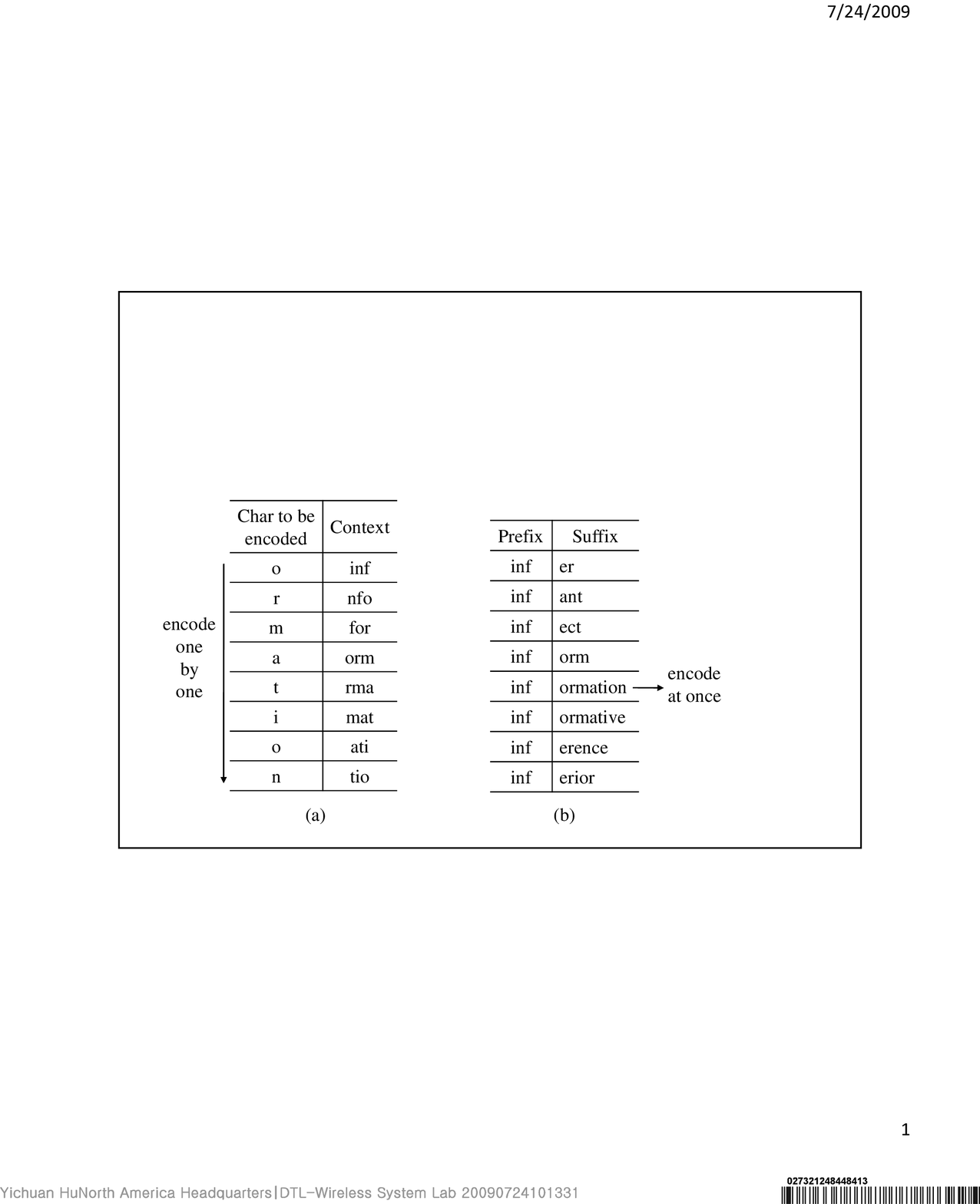}
   \vspace{0pt}
   \caption{Comparison of encoding the word "information" given that the first 3 characters "inf" have been encoded. (a) order-3 PPM encoder which processes characters one by one. (b) proposed variable-length prediction using a dictionary model that estimates rest characters of the word at once.}
   \label{fig_compare}
   \end{center}
   \vspace{0pt}
\end{figure}

First of all, a simple example is given to show how variable-length prediction can be achieved by a dictionary model. Suppose we are encoding a sequence of texts like this: ``...information...''. The first 3 characters ``inf'' have been encoded and characters to be encoded next are ``ormation...''. If we use order-$3$ PPM, characters are encoded one after another given its context, i.e. 3 preceding characters, as shown in Fig. \ref{fig_compare} (a). On the other hand, if we divide every word into two parts: a fixed-length prefix and a variable-length suffix, and suppose we have a dictionary that contains words with prefix ``inf'' as shown in Fig. \ref{fig_compare} (b). Instead of predicting characters one by one, we can predict suffixes in this dictionary at one time, provided that we know the prefix is ``inf''. Since the dictionary contains suffixes with different length, the prediction is variable-length.

Next, let's take a look at the advantage of using dictionary model. We still focus on the above example, and estimate how many bits are required to encode ``ormation''. For traditional order-$3$ PPM, assume 1.5 bits are needed on average for encoding one character, then we need $1.5 \times 8 = 12$ bits to encode ``ormation''. Notice that 1.5 bpc is not an unreasonable assumption because order-$3$ PPM usually achieves more than 2 bpc for text compression \cite{bell_89,moffat_tcom90}. On the other hand, as we can see from Fig. \ref{fig_compare} (b), there are 8 possible words that start with "inf" in the dictionary and ``information'' is one of them. Therefore, if we assign different indexes to these 8 words, we can easily encode ``ormation'' with as few as $\log_2{8} = 3$ bits, achieving 9 bits saving over order-$3$ PPM.

In practice, the dictionary model contains not only words but also their counts so that an arithmetic coder can be used to encode any word in the dictionary. Specifically, every dictionary $D$ includes three parts: a common prefix $P$, a list of strings $W_i$ that are suffixes of words with prefix $P$, and corresponding counts $C_i$. Given a prefix, we can find the associated dictionary model and then perform encoding/decoding. Moreover, different from context models in character-based PPM in which contexts from order-$n$ to order-$1$ are used, we only maintain dictionary models with fixed-length prefix (fixed context order). If the prefix is too short, each dictionary will contain a lot of words which is not good for efficient compression. On the other hand, if the prefix is too long, number of characters that can be predicted by dictionary model will be very small. In order to take full advantage of dictionary model, we choose the length of prefix as 3.

\subsection{Combining Dictionary Model and Context Model}

As we can see, the dictionary model works on word basis. By word we mean a sequence of consecutive English letters. We can always parse a text file into a sequence of alternating words and non-words. For non-words, they cannot be encoded by dictionary model, and even for the prefix of a word we need to encode character by character. Therefore, dictionary model is combined with context model used in traditional character-based PPM. At the beginning of encoding/decoding, both context model and dictionary model are empty, and texts are encoded on character basis. After a word is encoded, corresponding dictionary model is updated. Detailed encoding and decoding algorithms will be introduced in Section \ref{sec_details}.

Compared with the methods in \cite{horspool_dcc92} and \cite{skibinski_06}, the above framework has two advantages. First, no extra bits are required to indicate the switch from context model to dictionary model, because every time after the prefix of a word (3 consecutive English letters) is encoded or decoded by context model, encoder or decoder will automatically switch to dictionary model. Moreover, since dictionary models are constructed and updated during encoding/decoding, no pre-processing is required to build initial dictionaries.

\section{Algorithm Details}
\label{sec_details}

\subsection{Model Switching Using Finite State Machine (FSM)}
Any compression algorithms using more than one model face the problem of model switching \cite{volf_dcc98}. For example, in traditional PPM in which up to $n+1$ context models (order-$n$ to order-$0$) might be used when encoding a character, $escape$ code is sent as a signal to let decoder know switch from current order to lower order. In our case, we need a mechanism to control the switch between dictionary model and context model. We use finite state machine (FSM), and for both encoder and decoder there are 3 states: $S_0$, $S_1$ and $S_2$. The transition rules between states for encoding and decoding are different, which will be described next.

\subsection{Encoding Algorithm}

\begin{figure}[t]
   \begin{center}
   \includegraphics[width=0.56\columnwidth,keepaspectratio,clip]{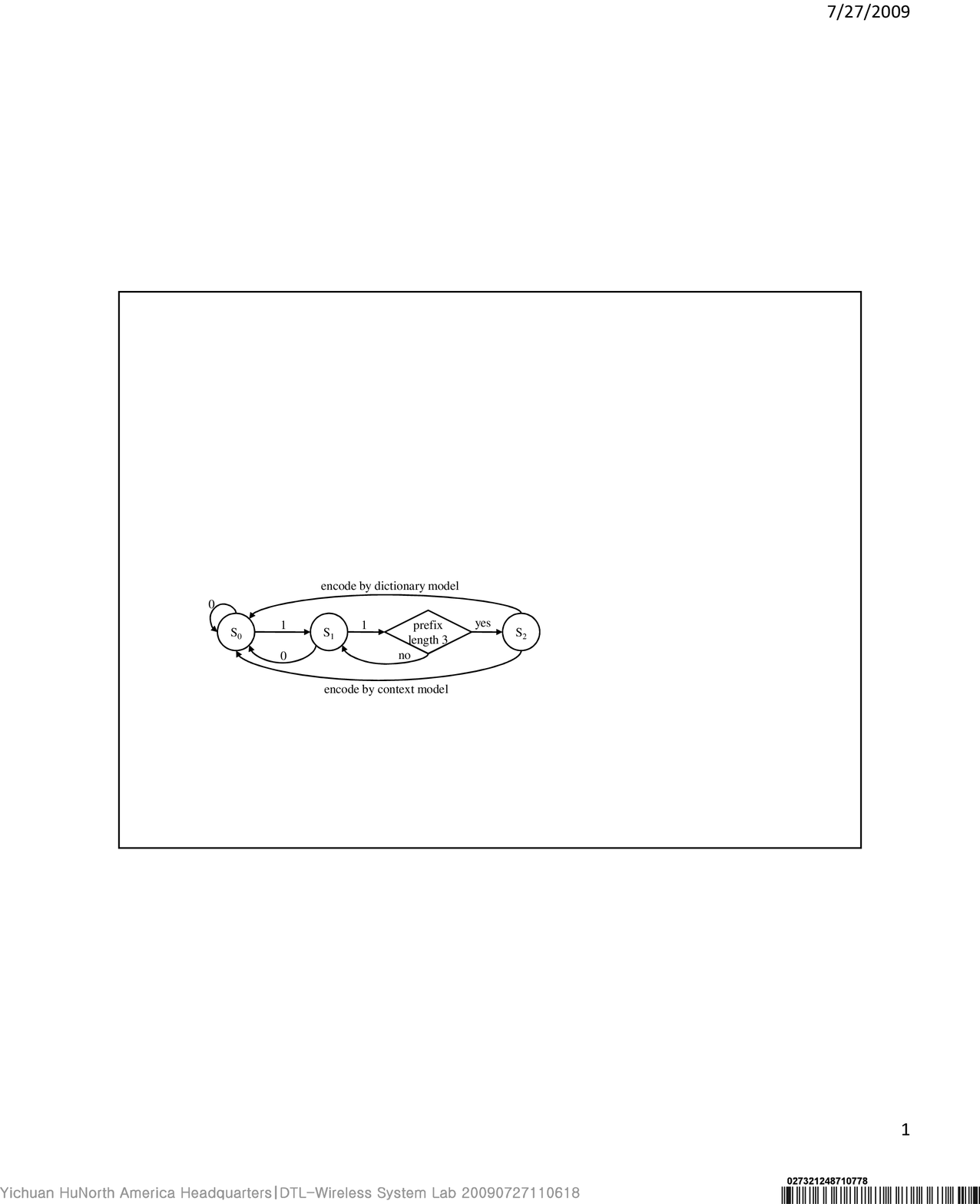}
   \vspace{0pt}
   \caption{Encoding algorithm using FSM. ``1'' and ``0'' denote the condition that next character to be encoded is an English or a non-English letter, respectively.}
   \label{fig_encoder}
   \end{center}
   \vspace{0pt}
\end{figure}

The encoding process starts at $S_0$, and the state transition rules are as follows:
\begin{itemize}
\item
At $S_0$: Encode the next character using context model. If it is an English letter, assign it to an empty string $P$ and move to $S_1$; otherwise stay at $S_0$.
\item
At $S_1$: Encode the next character using context model. If it is an English letter, append it to string $P$; otherwise, go back to $S_0$. If the length of $P$ reaches 3, move to $S_2$.
\item
At $S_2$: Read consecutive English letters, i.e. suffix string of current word, denoted as $W$. If a dictionary $D$ associated with prefix $P$ if found and $W$ exists in $D$, encode $W$ using dictionary model $D$. Otherwise, encode characters in $W$ one by one using context models (an $escape$ code should be encoded using dictionary model $D$ if $W$ cannot be found in $D$). Move to $S_0$.
\end{itemize}
The encoding state transition diagram is depicted in Fig. \ref{fig_encoder}. Pseudo code of the algorithm is provided below. On line 6 and 10, $English(c)$ is a function that checks character $c$ is an English letter or not. Due to limited space, ``\textbf{break}'' at the end of each ``\textbf{case}'' clause and the ``\textbf{default}'' clause are omitted.

\begin{algorithm}[H]
\label{alg_encoding}
\caption{VLPPM-Encoder}
$state \leftarrow S_0$\;
\While{$c \neq $ EOF}{
    encode next character $c$ using context model\;
    \Switch{$state$}{
    \Case{$S_0$}{
        \If{English($c$)}{
            $P \leftarrow c$\;
            $state \leftarrow S_1$\;
        }
    }
    \Case{$S_1$}{
        \lIf{English($c$)}{$P \leftarrow P + c$}\;
        \lElse{$state \leftarrow S_0$}\;
        \lIf{length($P$) = 3}{$state \leftarrow S_2$}\;
    }
    \Case{$S_2$}{
        read suffix $W$\;
        \If{find dictionary $D$ with prefix $P$}{
            \lIf{find $W$ in $D$}{encode $W$ using $D$}\;
            \Else{
                encode $escape$ code using $D$\;
                encode $W$ using context model\;
            }
        }
        \lElse{
            encode $W$ using context model\;
        }
        update dictionary model\;
        $state \leftarrow S_0$\;
    }

    }
}
\end{algorithm}

At state $S_2$, the probability of $escape$ code is calculated by
\begin{equation}
\label{eq_escape}
P_{escape} = \frac{1}{1+\sum_{W_i \in D}{C_i}},
\end{equation}
and the probability for suffix $W_i$ is
\begin{equation}
\label{eq_escape}
P_{W_i} = \frac{C_i}{1+\sum_{W_j \in D}{C_j}}.
\end{equation}

\subsection{Decoding Algorithm}

Similar to the encoding algorithm, the decoding algorithm also uses FSM with 3 states and starts at $S_0$, but with different state transition rules:
\begin{itemize}
\item
At $S_0$: Decode the next character using context model. If it is an English letter, assign it to an empty string $P$ and move to $S_1$; otherwise stay at $S_0$.
\item
At $S_1$: Decode the next character using context model. If it is an English letter, append it to string $P$; otherwise, go back to $S_0$. If the length of $P$ reaches 3, decode using dictionary model. If a string of characters are decoded, move to $S_0$; if an $escape$ code is decoded, move to $S_2$.
\item
At $S_2$: Decode characters one by one using context model until a non-English letter is decoded. Move to $S_0$.
\end{itemize}

\begin{algorithm}[H]
\label{alg_decoding}
\caption{VLPPM-Decoder}
$state \leftarrow S_0$\;
\While{$c \neq $ EOF}{
    decode $c$ using context model\;
    \Switch{$state$}{
    \Case{$S_0$}{
        \If{English($c$)}{
            $P \leftarrow c$\;
            $state \leftarrow S_1$\;
        }
    }
    \Case{$S_1$}{
        \lIf{English($c$)}{$P \leftarrow P + c$}\;
        \lElse{$state \leftarrow S_0$}\;
        \If{length($P$) = 3}{
            \If{find dictionary $D$ with prefix $P$}{
                decode $W$ using $D$\;
                \lIf{$W$ is $escape$ code}{$state \leftarrow S_2$}\;
                \lElse{update dictionary model}\;
            }
            \lElse{$state \leftarrow S_2$}\;
        }
    }
    \Case{$S_2$}{
        decode $c$ using context model\;
        \lIf{!English($c$)}{$state \leftarrow S_0$}\;
    }
    }
}
\end{algorithm}

\begin{figure}[t]
   \begin{center}
   \includegraphics[width=0.67\columnwidth,keepaspectratio,clip]{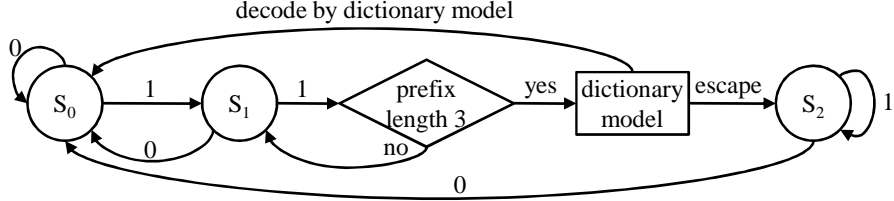}
   \vspace{0pt}
   \caption{Decoding algorithm using FSM. ``1'' and ``0'' denote the condition that the next decoded character is an English or a non-English letter, respectively.}
   \label{fig_decoder}
   \end{center}
   \vspace{0pt}
\end{figure}

\subsection{Exclusion}
\label{ssec_ex}
As we can see from the decoding algorithm, decoder automatically switches from context model to dictionary model once 3 consecutive English letters are decoded. By doing so, we don't need to waste any bits to indicate switches between dictionary model and context model. However, this leads to another problem: if a prefix in the dictionary is a word of length 3 (e.g., ``let'' can be either a word or a prefix of ``lettuce''), then every time this word occurs extra bits will be used by the dictionary model to encode an $escape$ code, resulting in performance degradation. To resolve this problem, we introduce an exclusion mechanism: after a word is encoded/decoded, if it is a prefix of a dictionary, this dictionary is discarded and this prefix is put in a ``blacklist'' for future reference. During encoding/decoding, if a prefix is in ``blacklist'', encoder/decoder skip dictionary model and use context model directly.

\section{Performance Evaluation}
\label{sec_results}
\subsection{Compression Efficiency}

In order to show the compression efficiency of the proposed algorithm, VLPPM encoder/decoder are implemented and tested on a large set of texts. Specifically, we implemented PPMC encoder/decoder as described in \cite{moffat_tcom90}, and further developed VLPPM based on PPMC. Text files from two popular data compression corpora, Calgary corpus \cite{bell_89} and Canterbury corpus \cite{canterbury}, are chosen as the data to be compressed. Compression ratio of VLPPM is presented in Table \ref{tab_compression} in terms of bits per character (bpc), and is compared with traditional PPM (PPMC). As we can see, using proposed VLPPM algorithm leads to considerable performance improvements: 16.3\% and 6.3\% gains are achieved over traditional PPM for order-$2$ and order-$3$, respectively. Moreover, although VLPPM implemented here is based on PPMC, it is applicable to any other character-based predictive compression schemes, such as all the variations of PPM \cite{moffat_tcom90,cleary_unbounded,bloom_ppmz,shkarin_dcc02}.

\begin{table}[t]
\caption{Compression Ratio Comparison of PPM and VLPPM}
\label{tab_compression} \centering
\begin{tabular}{c|c@{\hspace{5pt}}c@{\hspace{5pt}}c|c@{\hspace{5pt}}c@{\hspace{5pt}}c}
\hline
\hline
 & \multicolumn{3}{c|}{order-$2$} & \multicolumn{3}{c}{order-$3$}\\
\cline{2-7} File & PPM & VLPPM & Gain & PPM & VLPPM & Gain\\
\hline
\tt{bib} & 2.66 & 2.25 & 18.2\% & 2.12 & 1.99 & 6.5\%\\
\tt{book1} & 2.92 & 2.57 & 13.6\% & 2.48 & 2.36 & 5.1\%\\
\tt{book2} & 2.89 & 2.34 & 23.5\% & 2.27 & 2.07 & 9.7\%\\
\tt{news} & 3.26 & 2.87 & 13.6\% & 2.65 & 2.52 & 5.2\%\\
\tt{paper1} & 2.94 & 2.60 & 13.1\% & 2.50 & 2.40 & 4.2\%\\
\tt{paper2} & 2.89 & 2.52 & 14.7\% & 2.47 & 2.36 & 4.7\%\\
\tt{progc} & 2.91 & 2.71 & 7.4\% & 2.52 & 2.47 & 2.0\%\\
\tt{progl} & 2.40 & 2.13 & 12.7\% & 1.92 & 1.86 & 3.2\%\\
\tt{progp} & 2.29 & 2.05 & 11.7\% & 1.86 & 1.81 & 2.8\%\\
\tt{trans} & 2.38 & 2.08 & 14.4\% & 1.78 & 1.69 & 5.3\%\\
\tt{alice29.txt} & 2.72 & 2.41 & 12.9\% & 2.31 & 2.23 & 3.6\%\\
\tt{asyoulik.txt} & 2.81 & 2.58 & 8.9\% & 2.53 & 2.47 & 2.4\%\\
\tt{lcet10.txt} & 2.76 & 2.15 & 28.4\% & 2.19 & 1.97 & 11.2\%\\
\tt{plrabn12.txt} & 2.83 & 2.51 & 12.7\% & 2.44 & 2.33 & 4.7\%\\
\tt{cp.html} & 2.73 & 2.55 & 7.1\% & 2.38 & 2.34 & 1.7\%\\
\tt{fields.c} & 2.44 & 2.33 & 4.7\% & 2.18 & 2.15 & 1.4\%\\
\tt{grammar.lsp} & 2.72 & 2.60 & 4.6\% & 2.51 & 2.47 & 1.6\%\\
\tt{xargs.1} & 3.31 & 3.17 & 4.4\% & 3.08 & 3.05 & 1.0\%\\
\hline
Average & 2.86 & 2.46 & 16.3\% & 2.36 & 2.22 & 6.3\%\\
\hline
\hline
\end{tabular}
\end{table}

\begin{figure}[t]
   \begin{center}
   \includegraphics[width=0.6\columnwidth,keepaspectratio,clip]{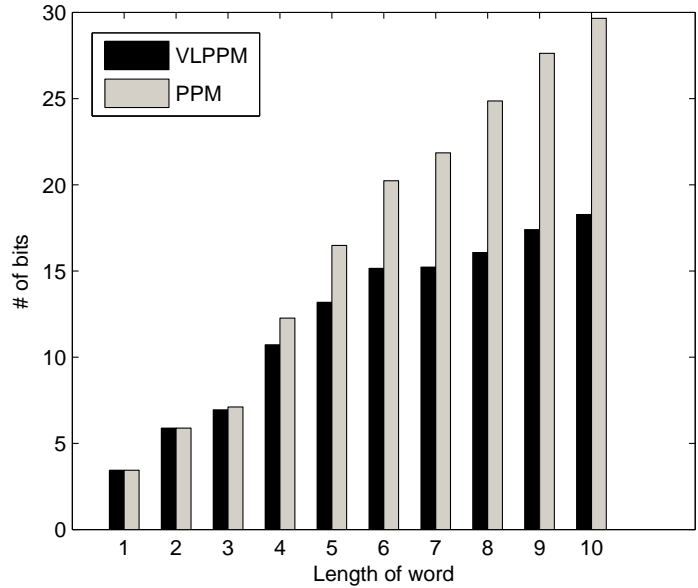}
   \vspace{0pt}
   \caption{Average number of bits per word when {\tt{lcet10.txt}} is compressed by order-$2$ PPM and VLPPM.}
   \label{fig_bit}
   \end{center}
   \vspace{0pt}
\end{figure}

In traditional PPM, characters are encoded one by one sequentially. As a result, the number of bits required to encode a word increases almost linearly with the word length. However, from the information theoretic point of view this is not the case because longer word doesn't necessarily contain more information. VLPPM, on the other hand, predicts several characters at once, achieving higher compression efficiency. This is illustrated in Fig. \ref{fig_bit}, in which the average number of bits required to encode words with different length is depicted for both PPM and VLPPM. As we can see, the number of bits increases almost linearly with word length when PPM is used, in accordance with our analysis above. Compared with PPM, VLPPM requires similar number of bits when word length is short, but much fewer bits when word length is long, and the longer the word is, the more bits can be saved.

\subsection{Computational Complexity}

Since VLPPM is based on PPM, it is natural to compare its complexity with that of PPM. Table \ref{tab_complexity} compares the time and memory consumption of VLPPM and PPM encoders for order-$2$ and order-$3$. In both cases, the results of VLPPM are presented in terms of fraction of time and memory compared with PPM. As we can see, the speed of VLPPM is comparable to PPM, and it is even faster than PPM at order-$2$, due to the variable-length prediction ability. Furthermore, although VLPPM uses dictionary models in addition to context models, the memory consumption only increases by a small percentage: 13\% and 8\% for order-$2$ and order-$3$, respectively. This is because VLPPM only maintains dictionaries with fixed-length prefix, i.e. prefix with length 3, which means only those words longer than 3 will be stored. Moreover, using the exclusion mechanism introduced in the Section \ref{ssec_ex} excludes certain words from being stored into dictionary which further reduces memory use.

\begin{table}[t]
\caption{Computational Complexity of PPM and VLPPM}
\label{tab_complexity} \centering
\begin{tabular}{c|cc|cc}
\hline
\hline
 & \multicolumn{2}{c|}{order-$2$} & \multicolumn{2}{c}{order-$3$}\\
\cline{2-5} & PPM & VLPPM & PPM & VLPPM\\
\hline
Time & 100\% & 98\% & 100\% & 108\%\\
Memory & 100\% & 113\% & 100\% & 108\%\\
\hline
\hline
\end{tabular}
\end{table}

\section{Conclusion}
\label{sec_con}
We have presented a text compression algorithm using variable-length prediction by partial matching (VLPPM). By introducing dictionary model which contains words with common prefix and combing it with context model used in traditional character-based PPM, the proposed method can predict one or more characters at once, further improving the compression efficiency without increasing computational complexity a lot. Moreover, the proposed method does not require any text preprocessing and can be applied to any other character-based predictive compression algorithms without increasing much computational complexity.

\end{document}